\documentclass[11pt]{amsart}
\usepackage{latexsym,amssymb,amsmath,amscd,amsthm}
\usepackage{amsfonts}
\numberwithin{equation}{section}
\theoremstyle{remark}

\newcommand{\bq}{\begin{equation}}
\newcommand{\bea}{\begin{array}}
\newcommand{\eea}{\end{array}}

\newcommand{\ga}{\alpha}
\newcommand{\gep}{\epsilon}
\newcommand{\gD}{\Delta}
\newcommand{\gl}{\lambda}

\newcommand{\gb}{\beta}

\newcommand{\mf}{\mathfrak}

\newcommand{\mc}{\mathcal}

\newcommand{\dg}{\dagger}

\newcommand{\go}{\omega}
\newcommand{\gO}{\Omega}
\newcommand{\gG}{\Gamma}
\newcommand{\gt}{\theta}
\newcommand{\gs}{\sigma}

\newcommand{\gag}{\gamma}
\newcommand{\gd}{\delta}
\newcommand{\pp}{\partial}

\newcommand{\tl}{\tilde}
\newcommand{\na}{\nabla}

\newcommand{\bl}{\blacklozenge}
\newcommand{\bs}{\blacksquare}

\newcommand{\gT}{\Theta}

\newcommand{\bgs}{\bigstar}

\newcommand{{\DDD}}{D\!\!\!\!\!\!-}

\title{SOME TOPICS IN THERMODYNAMICS AND QUANTUM MECHANICS}

\author{Robert Carroll\\University of Illinois, Urbana, IL 61801}

\date{November, 2012\thanks{email: rcarroll@math.uiuc.edu}}

\begin{document}

\bibliographystyle{plain}

\maketitle

\tableofcontents

\section{BACKGROUND}
\renewcommand{\theequation}{1.\arabic{equation}}
\setcounter{equation}{0}

The first theme here is to use fractional calculus (modified Riemann-Liouville - MRL (d'apres Jumarie
\cite{erly,jmre}) and q-deformed calculus (d'apres \cite{c002,hrmn}) to develop the Schr\"odinger
equation (SE) in various guises.  In particular we obtain fractional and q-defomed quantum potentials
(QP) and this context will have interaction with with e.g. q-Fisher information and various ideas of
entropy (e.g Renyi, Tsallis, etc.) for which we provide a very brief survey (cf. e.g. \cite{fred,frgy,frch,glts,gwhn,jbar,tsal,wgch}).
Hopefully this should help in 
developing thermodynamic (TD) connections with fractal situations and quantum mechanics (QM). 
We began this in \cite{cola}
where a fractional QP was outlined for a very elementary fractional derivative of V. Kobolev (cf.
\cite{kblv}) and in fact that treatment can be slightly refined, leading to the same conclusion
(cf. Section 2).  We do not review here the various fractional calculi (cf. \cite{erly,hrmn,jmre,
klst,klmz} but simply begin with MRL as in \cite{erly,jmre}.  Later we will expand this when discussing the
q-deformed calculus (as in \cite{hrmn}).  
\\[3mm]\indent
We go to the directly to the modified RL (MRL) integral of Jumarie following \cite{erly,jmre}.  Thus from \cite{erly}, pp. 1-69, with some stated and some unstated hypotheses,
define $({\bf 1A})\,\,F_w(h)f(x)=f(x+h)$ with fractional difference 
\bq\label{1.1}
\gD^{\ga}f(x)=(F_w-1)^{\ga}f(x)=\sum_0^{\infty}(-1)^k\left(
\begin{array}{c}
\ga\\
k\end{array}\right)f[x+(\ga-k)h]
\end{equation}
and fractional derivative
\bq\label{1.2}
f^{(\ga})(x)=lim_{h\to 0^{+}}\frac{\gD^{\ga}[f(x)-f(0)]}{h^{\ga}}
\end{equation}
While the RL involves
\bq\label{1.3}
f^{(\ga)}=\frac{1}{\gG(-\ga)}\int_0^x(x-\xi)^{-\ga-1}f(\xi)d\xi,\,\,(\ga<0)
\end{equation}
Jumarie uses the MRL $(0<\ga<1)$
\bq\label{1.4}
f^{(\ga)}(x)=[f^{\ga-1}(x)]'=\frac{1}{\gG(1-\ga)}\frac{d}{dx}\int_0^x(x-\xi)^{-\ga}[f(\xi)-f(0)]d\xi
\end{equation}
\bq\label{1.5}
f^{(\ga)}(x)=[f^{(\ga-n)}(x)]^{(n)}=\frac{1}{\gG(1-(\ga-n)}\frac{d^{n+1}}{dx^{n+1}}\int_0^x(x-\xi)^{-\ga-n)}
(f(\xi)-f(0))d\xi
\end{equation}
for $n\leq\ga<n+1$.
Laplace transforms are useful in proving some of the results in fractional calculus and we refer to 
\cite{erly,jmre} for details (some proofs are given here but generally we are simply stating results
in order to have a quick overview of the theory.
 If $f(x)$ is differentiable this is equivalent to the Liouville-Dzhrbashian-Caputo
definition
\bq\label{1.6}
f^{(\ga)}(x)=\frac{1}{\gG(1-\ga)}\int_0^x(x-\xi)^{-\ga}f'(\xi)d\xi\,\,(0<\ga<1)
\end{equation}
In general, to avoid non-commutativity problems one defines $({\bf 1B})\,\,D^{a+b}=D^{max(a,b)}D^{min
(a,b)}$.  A generalized Taylor expansion of fractional order is given by
\bq\label{1.7}
f(x+h)=\sum_0^{\infty}\frac{f^{\ga k}(x)}{(\ga k)!}h^{\ga k}\,\,(0<\ga\leq 1)
\end{equation}
where $f^{(\ga k)}(x)=D^{\ga}D^{\ga}\cdots D^{\ga}f(x)$ and $({\bf 1C})\,\,\gG(1+\ga k)=(\ga k)!$.
We note also that 
\bq\label{1.8}
F_w(h)=E_{\ga}(h^{\ga}D_x^{\ga});\,\,E_{\ga}(u)=\sum_0^{\infty}\frac{u^k}{(\ga k)!}
\end{equation}

\indent
The equation (1.7) provides the useful relation $({\bf 1D})\,\,d^{\ga}f=\gG(1+\ga)df,\,\,0<\ga<1$
or equivalently $({\bf 1E})\,\,\gD^{\ga}f\simeq \ga!\gD f$ which holds for non-differentiable functions
only.  One can also write from this
\bq\label{1.9}
D^{n+\gt}x^{\gag}=\gG(\gag+1)\gG^{-1}(\gag+1-n-\gt)x^{\gag-n-\gt}\,\,(0<\gt<1);
\end{equation}
$$(u(x)v(x))^{(\ga)}=u^{(\ga)}(x)v(x)+u(x)v^{(\ga)}(x);\,\,[f(u(x))]^{(\ga)}=f_u^{(\ga)}(u)(u'_x)^{\ga}$$
In this spirit, if both u and v are non-differentiable, one has $\ga!d(uv)=ud^{\ga}v+vd^{\ga}u$ and 
(1.9C) is a consequence of 
\bq\label{1.10}
\frac{d^{\ga}f(u)}{dx^{\ga}}=\frac{d^{\ga}f}{du^{\ga}}\left(\frac{du}{dx}\right)^{\ga}
\end{equation}
Note here that 
\bq\label{1.11}
d^{\ga}f=\ga! df\iff f(u)\,\,is\,\,non-differentiable
\end{equation}
Further we write
\bq\label{1.12}
\frac{d^{\ga}u}{du^{\ga}}=\frac{1}{(1-\ga)!}u^{1-\ga}\Rightarrow
\end{equation}
$$(du)^{\ga}=(1-\ga)u^{\ga-1}d^{\ga}u\Rightarrow
(u'_x(x))^{\ga}=(1-\ga)!u^{\ga-1}u_x^{(\ga)}(x)$$
Note that $u^{(\ga)}_x(x)$ may exist while $u'(x)$ is undefined but if $u'(x)$ exists then 
$u^{(\ga)}(x)$ exists with relations as in (1.12).  If $f(u)$ is differentiable and $u(x)$ is differentiable then
\bq\label{1.13}
(f[u(x)])^{(\ga)}=\frac{1}{(1-\ga)!}[f(u)]^{1-\ga}\left(\frac{df}{du}\right)^{\ga}\left(\frac{du}{dx}\right)^{\ga}
\end{equation}
If both $f(u)$ and $u(x)$ have fractional derivatives then 
\bq\label{1.14}
(f[u(x)])^{(\ga)}=(1-\ga)!u^{\ga-1}(f(u))^{1-\ga}f_u^{(\ga)}(u)u_x^{(\ga)}(x)
\end{equation}
To see this write
\bq\label{1.15}
\frac{d^{\ga}f(u(x))}{dx^{\ga}}=\frac{d^uf}{d^{\ga}u}\frac{d^{\ga}u}{dx^{\ga}}=\frac{d^{\ga}f}{du^{\ga}}\frac{du^{\ga}}{d^{\ga}u}\frac{d^{\ga}u}{dx^{\ga}}
\end{equation}
and use (1.12).  The integration with respect to $(dx)^{\ga}$ is defined as the solution of the fractional
equation $({\bf 1F})\,\,dy=f(x)(dx)^{\ga},\,\,x\geq 0,\,\,y(0)=0,\,\,0<\ga<1$ which is provided via the 
equation
\bq\label{1.16}
y=\int_0^xf(\xi)(d\xi)^{\ga}=\ga\int_0^x(x-\xi)^{\ga-1}f(\xi)d\xi\,\,\,(0<\ga\leq 1)
\end{equation}
More generally 
\bq\label{1.17}
\int_a^xf(\xi)(d\xi)^{\ga}=\ga\int_a^x(x-\xi)^{\ga-1}f(\xi-a)d\xi,\,\,\,(0<\ga\leq 1)
\end{equation}
and there is an integration by parts formula
\bq\label{1.18}
\int_a^bu^{(\ga)}(x)v(x)(dx)^{\ga}=\ga![u(x)v(x)]_a^b-\int_a^bu(x)v^{(\ga)}(x)(dx)^{\ga}
\end{equation}
If $y=g(x)$ is a non-decreasing differentiable function then 
\bq\label{1.19}
\int f(y)(dy)^{\ga}=\int f(g(x))(g'(x))^{\ga}(dx)^{\ga}\,\,0<\ga<1
\end{equation}
and when $g(x)$ has a positive fractional derivative of order $\gb,\,\,0<\ga,\,\gb<1$ there results
\bq\label{1/20}
\int f(y)(dt)^{\ga}=\gG^{-\ga}(1+\gb)\int f(g(x)(g^{(\gb)}(x))^{\ga}(dx)^{\ga\gb}
\end{equation}
A few special formulas are obtained by seting $f(x)=x^{\gag}$ in (4.29) to get
\bq\label{1.21}
\int_0^x\xi^{\gag}(d\xi)^{\ga}=\frac{\gG(\ga+1)\gG(\gag+1)}{\gG(\ga+\gag+1)}x^{\ga+\gag}
\end{equation}
One expects a generalized Dirac delta function to give $({\bf 1G})\,\,\int \gd(\xi)(d\xi)^{\ga}=\ga x^{\ga-1}$
and using (1.9) the fractional derivative of this delta function can be defined via
$({\bf 1H})\,\,\int\gd^{(\ga)}(\xi)f(\xi)(d\xi)^{\ga}=-\int \gd(\xi)f^{(\ga)}(\xi)(d\xi)^{\ga}\,\,(0<\ga\leq 1)$
and (1.17) gives then $({\bf 1I})\,\,\int \gd^{(\ga)}(\xi)f(\xi)(d\xi)^{\ga}=-\ga x^{\ga-1}f^{(\ga)}(0)\,\,\,
(0<\ga\leq 1)$.  The relation between fractional integral and derivative is
\bq\label{1.22}
\frac{d^{\ga}}{dx^{\ga}}\int_0^xf(\xi)(d\xi)^{\ga}=\gG)1+\ga)f(x)=\ga! f(x);
\end{equation}
$$\frac{d}{dx^{\ga}}\int_0^{u(x)}
f(\xi)(d\xi)^{\ga}=\ga! f(u(x))(u'(x))^{\ga}$$
This leads to the useful result $(\bgs)\,\,y=(1/\ga!)\int_0^xy^{(\ga)}(\xi)(d\xi)^{\ga}$.
\\[3mm]\indent
Now for coarse graining and fractal space it appears that $\ga$ should be the box (or perhaps
Hausdorff) dimension of the space determined via $({\bf 1J})\,\,d=lim_{\gep\to 0^{+}}[-(log N(\gep)/log(\gep)]$ where $N(\gep)$ is the number of $\epsilon$ balls needed to cover the curve.  Thus $N\sim (L/\gep^{\ga})$
and $({\bf 1K})\,\,d=lim_{\gep\to 0}[-log(\gep^{\ga}/log(\gep)]$.  In this framework the velocity $\dot{x}_{\ga}
(t)$ is defined by $({\bf 1L})\,\,u_{\ga}(t)=\dot{x}_{\ga}(t)=(dx)^{\ga}/dt;\,\,dx>0,\,\,0<\ga<1$.  
For coarse graining in both time and space we have $(0<\gb<1,\,\,(dt)^{\gb}>dt)$
\bq\label{1.23}
v_{\gb}(t)=x^{(\gb)}(t)=\frac{d^{\gb}x}{(dt)^{\gb}}=\gb!\frac{dx}{(dt)^{\gb}}
\end{equation}
This leads to a time-space coarse graining
\bq\label{1.24}
\dot{x}_{\ga,\gb}(t)=\frac{(dx)^{\ga}}{(dt)^{\gb}}=\left[\frac{(dx)^{\ga/\gb}}{dt}\right]^{\gb}=
(\dot{x}_{\ga/\gb})^{\gb}\,\,(0<\ga <\gb<1)
\end{equation}
Some preliminary results are e.g.
\begin{enumerate}
\item
Given $y=f(x)$ and $x=g(y)$ one has $({\bf 1M})\,\,y^{(\ga)}(x)x^{(\ga)}(y)=[(1-\ga)!]^{-2}(xy)^{1-\ga}$.
This follows from 
\bq\label{1.25}
y^{(\ga)}(x)x^{(\ga)}(y)=\left(\frac{d^{\ga}y}{dx^{\ga}}\right)\left(\frac{d^{\ga}x}{dy^{\ga}}\right)=\left(\frac{d^{\ga}y}{dy^{\ga}}\right)\left(\frac{d^{\ga}x}{dx^{\ga}}\right)
\end{equation}
and $({\bf 1N})\,\,D^{\ga}x^{\gag}=\gG(\gag+1)\gG^{-1}(\gag+1-\ga)x^{\gag-\ga}$ which implies
$({\bf 1O})\,\,(1-\ga)!d^{\ga}x=x^{1-\ga}(dx)^{\ga}$.  If now we assume that $x(t)$ is a function of time which is discontinuous in such a manner that one can write $d^{\ga}t=\ga!dt$
then ({\bf 1L}) implies
\bq\label{1.26}
\dot{x}_{\ga}(t)=\frac{\ga!(dx)^{\ga}}{\ga!dt}=\ga!\frac{(dx)^{\ga}}{d^{\ga}t}=\ga![t^{(\ga)}(x)]^{-1}
\end{equation}
Putting ({\bf 1M}) into (1.26) yields
\bq\label{1.27}
\dot{x}_{\ga}(t)=\ga!((1-\ga)!)^2(xt)^{\ga-1}x^{(\ga)}(t)=\rho(\ga)(xt)^{\ga-1}x^{(\ga)}(t)
\end{equation}
with $\rho(\ga)=\ga![(1-\ga)!]^2$.
\end{enumerate}
The Mittag-Leffler function is defined via $({\bf 1P})\,\,E_{\ga}(t^{\ga})=\sum_0^{\infty}[t^{\ga k}/(\ga k)!]$
where $(\ga k)!=\gG(1+\ga k)$.  It satisfies the equation 
$({\bf 1Q})\,\,D^{\ga}E_{\ga}(t^{\ga})=E_{\ga}(t^{\ga})$ and further
\bq\label{1.28}
D^{\ga}K=0\,\,(K=constant);\,\,D^{\ga}x^{\ga k}=\frac{(\ga k)!}{(\ga k-\ga)!}x^{\ga(k-1)}
\end{equation}
$$f(x+h)= E_{\ga}[h^{\ga}D_x^{\ga}f(x)]$$
One can state in particular
\\[3mm]\indent
{\bf PROPOSITION 1.1.}
Given two real functions $f,g:\,{\bf R}\to {\bf R}$ of the form $f\to f(u)$ and $x\to u(x)$ then 
\bq\label{1.29}
\frac{d^{\ga}f(u(x))}{dx^{\ga}}=\frac{d^{\ga}f(u)}{du^{\ga}}\left(\frac{du}{dx}\right)^{\ga};
\end{equation}
$$f_x^{(\ga)}(u(x))=(1-\ga)!u^{\ga-1}f_u^{(\ga)}(u)u_x^{(\ga)}(x)$$
\indent
{Proof:}
First note that (1.29A) follows in writing 
\bq\label{1.30}
\frac{d^{\ga}f(u(x))}{dx^{\ga}}=\frac{d^{\ga}f(u)}{du^{\ga}}\frac{(du)^{\ga}}{(dx)^{\ga}}=\frac{d^{\ga}f(u)}{du^{\ga}}(u')^{\ga}
\end{equation}
Then putting (1.12) into (1.30) gves
\bq\label{1.31}
\frac{d^{\ga}f}{dx^{\ga}}=\frac{d^{\ga}f(u)}{du^{\ga}}(1-\ga)!u^{\ga-1}\frac{d^{\ga}u}{dx^{\ga}}
\end{equation}
Note that if a function has a derivative then it has an $\ga$ derivative for any $\ga\,\,(0<\ga<1)$.
As an example note that for $f(x)=E_{\ga}(\gl log^{\ga}(x)$
\bq\label{1.32}
\frac{d^{\ga}f(x)}{dx^{\ga}}=\frac{d^{\ga}E_{\ga}(\gl log^{\ga}(x))}{(d(log(x))^{\ga}}\left(\frac{d(log(x)}{dx}\right)^{\ga}=\gl x^{-\ga}E_{\ga}(\gl log^{\ga}(x)
\end{equation}
Another example involves $y=y(x),\,\,x=g(y),\,\,y=f(g(y))$.  Then
\bq\label{1.33}
y^{(\ga)}(x)x^{(\ga)}(y)=\frac{d^{\ga}y}{dx^{\ga}}\frac{d^{\ga}x}{(dy)^{\ga}}=\frac{d^{\ga}y}{dy^{\ga}}
\frac{d^{\ga}x}{dx^{\ga}}
\end{equation}
Taking account of (1.9) there results 
\bq\label{1.34}
y^{(\ga)}(x)x^{\ga}(y)=(1-\ga)!)^{-2}(xy)^{1-\ga}
\end{equation}
As a variation consider $y=E_{\ga}(x)$ and $x^{\ga}=Log_{\ga}y$ where
$({\bf 1R})\,\,y^{\ga}(x)=E_{\ga}(x^{\ga})$ and (1.34) yields then
\bq\label{1.35}
E_{\ga}(x^{\ga})x^{(\ga)}(y)=((1-\ga)!)^{-2}(xy)^{1-\ga}
\end{equation}
Therefore
\bq\label{1.36}
x^{(\ga)}(y)=D^{\ga}[(Log_{\ga}(y))^{1/\ga}]=\frac{1}{(1-\ga)!)^2}\frac{(xy)^{1-\ga}}{E_{\ga}(x^{\ga})}
\end{equation}
Consequently
\bq\label{1.37}
D^{\ga}[Log_{\ga}(y))^{1/\ga}]=\frac{1}{((1-\ga)!)^2}\frac{y^{1-\ga}}{y}[(Log_{\ga}(y)]^{\frac{1-\ga}{\ga}}
\end{equation}
\\[3mm]\indent
A function $x(t)$ is said to be self-similar with Hurst exponent $H>0$ whenever
$({\bf 1S})\,\,x(at)\propto a^Hx(t)\Rightarrow x(t)\simeq t^Hx(1)$.
(1.18) implies also $({\bf 1T})\,\,x(t)-x(0)\simeq bt^{\ga}$ for small $t$ and therefore
$x(at)-x(0)\simeq a^{\ga}(x(t)-x(0))$.  This means that if $x(t)$ is $\ga$ differentiable
at $t=0$ then $x(t)x(0)$ is locally self similar with Hurst exponent $\ga$ and 
there is also a kind of converse.  In any event self similarity and Hurst 
exponents are intimately connected with fractional calculus and since self similarity and
H\"older exponents are characteristic of fractals we can see how the fractal-fractional
connection arises (cf. also \cite{ccgn}).  
We add now a few formulas from \cite{jmre} (especially from papers in 2009 and 2012 - with
possible repetitions from above):
\bq\label{1.38}
\int\frac{d^{\ga}x}{x}=Log_{\ga}\left(\frac{x}{C}\right);\,\,x=E_{\ga}(Log_{\ga}(x));\,\,[x/C]>0)
\end{equation}
Further
\begin{enumerate}
\item
$Log_{\ga}(x^y)=y^{\ga}Log_{\ga}(x);\,\,E_{\ga}(x^{\ga}y^{\ga})=E_{\ga}(y^{\ga})^x$
\item
$(Log_{\ga}(uv))^{1/\ga}=(Log_{\ga}u)^{1/\ga}+(Log_{\ga}v)^{1/\ga};\,\,(dx)^{\ga}=(\ga!)^{-1}d^{\ga}(x^{\ga})$
\item
$E_{\ga}(\gl(x+y)^{\ga})=E_{\ga}(\gl x^{\ga})+E_{\ga}(\gl y^{\ga});\,\,\ga!(1-\ga)!dx=x^{1-\ga}(dx)^{\ga}$
\item
For MRL but not RL one has
$$D^{\ga}E_{\ga}(\gl x^{\ga})=\gl E_{\ga}(\gl x^{\ga});\,\,D^{\ga}E_{\ga}(u^{\ga}(x))=
E_{\ga}(u^{(\ga)})(u'(x)^{\ga})$$
\end{enumerate}
Another collection of formulas from \cite{jmre} (APL, 22 (2009), 378-385) is given 
(with possible repetitions) via ($\gG_{\ga}(n+1)=(\ga!)^n n!$)
\bq\label{1.39}
D^{\ga}E_{\ga}(\gl x)=\gl\ga^{-\ga}x^{1-\ga}E_{\ga}(\gl x);
\,\,\gG_{\ga}(x)=(\ga!)^{-1}\int_0^{\infty}E_{\ga}(-t^{\ga})t^{\ga-1)\ga}(dt)^{\ga};
\end{equation}
$$\int (dx)^{\ga}=\gG^{-1}(2-\ga)\int x^{1-\ga}(dx)^{\ga};\,\,\gG_{\ga}(x+1)=(\ga!)x\gG_{\ga}(x);$$
$$\frac{d^{\ga}Log_{\ga}(x)}{dx^{\ga}}=\frac{\ga!}{(1-\ga)!}\frac{1}{x^{\ga}};\,\,B_{\ga}(x,y)=\int_0^1(1-t)^{(x-1)
\ga}t^{(y-1)\ga}(dt)^{\ga}$$
$$\int f^{(\ga})(x)(dx)^{\ga}=\int d^{\ga}f=\ga!f(x);\,\,B_{\ga}(x,y)=\frac{\gG_{\ga}(x)\gG_{\ga}(y)}{\gG_{\ga}
(x+y)}$$
$$Log_{\ga}(x)=\gG^{-1}(2-\ga)\int_0^x\left(\frac{d\xi}{\xi}\right)^{\ga};\,\,D^{\ga}g(u(x),x)=g'_uu^{(\ga)}(x)+\gG^{-2}(2-\ga)x^{1-\ga}g'_x$$
$$x=E_{\ga}\left(\gG^{-1}(2-\ga)\int_0^x\left(\frac{d\xi}{\xi}\right)^{\ga}\right);\,\,
\int\frac{(dx)^{\ga}}{x}=\frac{x^{\ga-1}}{(\ga-1)\gG(2-\ga)};$$
$$\int_{{\bf R}}E_{\ga}\left(-\left(\frac{x^2}{2\gs^2}\right)^{\ga}\right)(dx)^{\ga}=(2\gs)^{\ga}\sqrt{\frac{(\ga!)^3}{(2\ga)!}}\pi^{\gs/2}$$

\section{FRACTIONAL SCHR\"ODINGER EQUATION}
\renewcommand{\theequation}{2.\arabic{equation}}
\setcounter{equation}{0}

We go here to \cite{jmre} (CSF, 41 (2009), 1590-1604) and write first some useful formulas:
\bq\label{2.1}
v_{\gb}(t)=x^{(\gb)}(t)=\frac{d^{\gb}x}{(dt)^{\gb}}=\gb!\frac{dx}{(dt)^{\gb}}
\end{equation}
\bq\label{2.2}
y^{(\ga)}(x)x^{(\ga)}(y)=\left(\frac{d^{\ga}y}{dx^{\ga}}\right)\left(\frac{d^{\ga}x}{dy^{\ga}}\right)=\left(\frac{d^{\ga}y}{dy^{\ga}}\right)\left(\frac{d^{\ga}x}{dx^{\ga}}\right)
\end{equation}
\bq\label{2.3}
u_{\ga}=\dot{x}_{\ga}(t)=\frac{\ga!(dx)^{\ga}}{\ga!dt}=\ga!\frac{(dx)^{\ga}}{d^{\ga}t}=\ga![t^{(\ga)}(x)]^{-1}]
\end{equation}
From ({\bf 1M}) and (2.3) we get then
\bq\label{2.4}
u_{\ga}(t)=\ga!((1-\ga)!)^2(xt)^{\ga-1}x^{(\ga)}(t)=\rho(\ga)(xt)^{\ga-1}x^{(\ga)}(t)
\end{equation}

\indent
For the Schr\"odinger equation (SE) we can look at a Lagrangian and action ($q\sim x$)
\bq\label{2.5}
S=\int_{t_1}^{t_2}L(q,u_{\ga},\tau)(d\tau)^{\ga};\,\,L=K(q,u_{\ga},t)-V(q);\,\,K=\frac{1}{2}\sum_{ij}(u_{\ga})_i(u_{\ga})_j
\end{equation}
where $V(q)$ is a potential energy term.  Then (assuming suitable differentiability)
\bq\label{2.6}
\gd S=\int_{t_1}^{t_2}\left[\frac{\pp L}{\pp q}\gd q+\frac{\pp L}{\pp u_{\ga}}\frac{\pp u_{\ga}}{\pp q}\gd q+\frac
{\pp L}{\pp u_{\ga}}\frac{\pp u_{\ga}}{\pp q^{\ga}}(\gd q)^{\ga}\right](dt)^{\ga}
\end{equation}
In order to convert the integral involving $(\gd q)^{\ga}$ to an integral in terms of $\gd q$ one uses the formula
\bq\label{2.7}
\int (fg)^{(\ga)}(dt)^{\ga}=\int f^{(\ga)}(dt)^{\ga}+\int fg^{(\ga)}(dt)^{\ga}
\end{equation}
leading to the dynamical equation
\bq\label{2.8}
\frac{\pp L}{\pp q}-\frac{d^{\ga}}{dt^{\ga}}\left(\frac{\pp L}{\pp u_{\ga}}\frac{\pp u_{\ga}}{\pp q^{(\ga)}}\right)+F=0;\,\,
\frac{\pp L}{\pp q}-\frac{d^{\ga}}{dt^{\ga}}\left(\frac{\pp L}{\pp q^{(\ga)}}\right)+F=0
\end{equation}
where F denotes an external applied force.  Then write in Hamiltonian form
\bq\label{2.9}
H\sum p_iq_i^{(\ga)}-L(q,u_{\ga}(q,q^{(\ga)},t),t);\,\,q^{(\ga)}(t)=\frac{\pp H}{\pp p};\,\,p^{(\ga)}(t)=-\frac{\pp H}{\pp q}
\end{equation}
which are strictly equivalent to (2.8).
\\[3mm]\indent
Now a probability density of fractional order is introduced as follows.  Let X be a random variable
on $[a,b]$ and $p_{\ga}(x)$ be a positive function on $[a,b]$.  Then X is referred to as a random
variable of fractional order $\ga,\,\,0<\ga<1$ with probability $p_{\ga}(x)$ whenever 
\bq\label{2.10}
F(x',x)=Pr(x'<X<x)=\frac{1}{\gG(1+\ga)}\int_{x'}^xp_{\ga}(\xi)(d\xi)^{\ga};\,\,F(a,b)=1
\end{equation} 
According to (2.10) and
\bq\label{2.11}
dy=(f(x)(dx)^{\ga}\Rightarrow
y=\int_0^xf(\xi)(d\xi)^{\ga}=\ga\int_0^x(x-\xi)^{\ga-1}f(\xi)d\xi;\,\,(0<\ga\leq 1)
\end{equation}
$$\Rightarrow \frac{d}{dx^{\ga}}\int_0^xf(\xi)(d\xi)^{\ga}=\ga!f(x)$$
if we denote by $p(x)$ the corresponding probability density of X we should have the identity
\bq\label{2.12}
p(x)=\gG^{-1}(\ga)(a-x)^{\ga-1}p_{\ga}(x),\,\,(x\leq a)
\end{equation}
In other words the fractional probability density $p_{\ga}(x)$ can be thought of as a family of standard probability density functions $p(x)$.  Then using (2.10) there results
\bq\label{2.13}
\frac{\pp^{\ga}F(x',x)}{\pp x^{\ga}}=p_{\ga}(x)
\end{equation}

\indent
Now the Hamiltonian function associated with a particle in a force field determined by the potential function
$V(x,t)$ is $({\bf 2A})\,\,H=(1/2m)p^2+V(x,t)$ with $p=mv$ and for the SE one takes $({\bf 2B})\,\,
p\to -i\hbar\na,\,\,p^2\to -\hbar^2\gD$ so that $({\bf 2C})\,\,i\hbar\psi_t=-(\hbar^2/2m)\psi_{xx}+V(x,t)$.
Classically 
\bq\label{2.14}
E=(1/2m)p^2+V,\,\, E\to -i\hbar\pp_t,\,\, i\hbar\psi_t=-(\hbar^2/2m)\psi_{xx}+V\psi
\end{equation}
Now write the energy as $({\bf 2D})\,\,E=(1/2m)\rho^2(\ga)(xt)^{2(\ga-1)}p_{\ga}^2$,
based on $p_{\ga}\sim mx^{(\ga)}\sim-i\hbar\na$, with velocity $({\bf 2E})\,\,
u_{\ga}=(dx)^{\ga}/dt=\rho(\ga)(xt)^{\ga-1}x^{(\ga)}$ as in (2.4).
Then with $E\to i\hbar\pp_t^{(\ga)}$ one obtains the SE
\bq\label{2.15}
i\hbar\psi_t^{(\ga)}=-\frac{\hbar^2}{2m}\rho^2(\ga)(xt)^{2(\ga-1)}\psi_{xx}+V\psi
\end{equation}
This last step is somewhat heuristic and it is justified in \cite{jmre} by 
constructing an example where it can hold.  We would prefer however to consider Sch\"odinger equations
written directly in terms of fractional derivatives, e.g. 
\bq\label{2.16}
i\hbar D_t^{\gb}\psi=-\frac{\hbar^2}{2m}D_x^{2\ga}\psi+V\psi
\end{equation}
which facilitate the introduction of quantum potentials (QP).
\\[3mm]\indent
In the case of (2.15) one could imagine a QP of the form 
\bq\label{2.17}
\hat{{\mf Q}}=-\frac{\hbar^2}{2m}\rho^2(\ga)(xt)^{2(\ga-1)}\frac{R_{xx}}{R}
\end{equation}
and the entrance of time here and in (2.15) suggests perhaps memory effects in the fractional
context (cf. \cite{hrmn,lask,trsv}).
In \cite{cola} we worked with a fractional SE $({\bf 2F})\,\,i\hbar\pp_t\psi=(\hbar^2/2m)d^2_{\ga}\psi+V(x)\psi$ where $d_{\ga}$ was a very special fractional derivative $({\bf 2G})\,\,d_{\ga}z^{\gb}=\gb z^{\gb-\ga}$ and a space of Puiseux functions was used (cf. \cite{cola,kblv}).
A function $\tl{E}_{\ga}$ (resembling $E_{\ga}$ in its properties) was used with a wave function $\psi=R\tl{E}_{\ga}(iS/\hbar)$ to solve  ({\bf 2F}) and
determine a quantum potential (QP) (cf. Remark 2.1 below)
\bq\label{2.18}
{\mf Q}_{\ga}=-\frac{\hbar^2d^2_{\ga}R}{2mR}
\end{equation}
Here it might be possible to find a similar construction in the MRL theory and we note in passing that a 
possibly macro
framework for the SE and QP arises in the Nottale theory (\cite{nott,notl}).   
The time variable will be assumed 
classical here and $D^{2\ga}=D^{\ga}D^{\ga}$ for simplicity ($(R,S)$ depend on $(x,t)$).
\\[3mm]\indent
{\bf REMARK 2.1.}
The function $\tl{E}_{\ga}$ of \cite{kblv} can be written as
\bq\label{2.19}
\tl{E}_{\ga}(z)=\sum_0^{\infty}\frac{\left(z^{\ga}/\ga)\right)^k}{\gG(k+1)}=exp\left(\frac{z^{\ga}}{\ga}\right)
\end{equation}
This does not seem to be in the general Mittag-Leffler class (cf. \cite{erly,hrmn,jmre,klst}) however.
In any event in \cite{cola} we used $({\bf 2H})\,\,\tl{E}_{\ga}=\sum_0^{\infty}(z^{\ga}/\ga)^k/\gG(k+1)\sim
exp[z^{\ga}/\ga)]$ so that
\bq\label{2.20}
d_{\ga}\left(\frac{z^{\ga}}{\ga}\right)^k=d_{\ga}\left(\frac{z^{\ga k}}{\ga^k}\right)=\ga^{-k}\ga kz^{\ga k-\ga}=\ga^{1-k}kz^{\ga(k-1)}
\end{equation}
This means that
\bq\label{2.21}
d_{\ga}\tl{E}_{\ga}(z)=\sum_1^{\infty}\frac{\ga^{1-k}}{(k-1)!}z^{\ga(k-1)}
=\sum_0^{\infty}\left(\frac{z}{\ga}\right)^n\frac{1}{n!}=\tl{E}_{\ga}(z)
\end{equation}
We note here an oversight in \cite{cola} (math-ph 1206.0900) where we used an
analogue of $E_{\ga}$, i.e. $\tl{E}_{\ga}(z)=exp[z^{\ga}/\ga)]=\sum_0^{\infty}(z^{\ga}/\ga)^k/k!$.
With the Kobelev derivative in \cite{kblv} given by $d_{\ga}z^{\gag}=\gag z^{\gag-\ga}$ one has
$d_{\ga}\tl{E}_{\ga}=\tl{E}_{\ga}$ and there is a chain rule for the $d_{\ga}$ calculus.  In \cite{cola}
we wrote $\tl{E}_{\ga}(i\hat{S})$ for solution of a Schr\"odinger equation (= SE) $({\bf 2I})\,\,i\hbar\psi_t=-[(\hbar^2)(2m)]\gD\psi+(V/2m)\psi$ and overlooked some features $(\hat{S}=S/\hbar$ is used).
In particular in using $\tl{E}_{\ga}$ we should
have emphasized $i\hat{S}\sim z^{\ga}/\ga$ and the calculations are then somewhat different; however 
the result is again (2.18).  
Thus if we take $z^{\ga}/\ga=i\hat{S}$ with $S=S(x,t)$ there results from \cite{kblv} $({\bf 2J})\,\,d_{\ga}^z\tl{E}(z)=\tl{E}(z)$ and e.g. $({\bf 2K})\,\,\pp_t\tl{E}_{\ga}(z)=\tl{E}_{\ga}(z)\pp_tz$ where
\bq\label{2.22}
i\hat{S}_t=\frac{1}{\ga}\ga z^{\ga-1}\pp_tz \Rightarrow \frac{\hat{S}_t}{\hat{S}}=
\frac{z_t}{z}
\end{equation}
Also
\bq\label{2.23}
id_{\ga}^x\hat{S}=z^{\ga-1}d^x_{\ga}z=z^{\ga-1}z_{\ga}^x;
\end{equation}
$$id^x_{2\ga}\hat{S}=d_{\ga}^x(z^{\ga-1}z_{\ga}^x)=
(\ga-1)z^{\ga-2}(z_{\ga}^x)^2+z^{\ga-1}z_{2\ga}^x$$
By analogy with the classical SE with $d_{\ga}^x\sim\pp_x$ we can posit
\bq\label{2.24}
i\hbar(R_t\tl{E}_{\ga}+R\tl{E}_{\ga}z_t)=-\frac{\hbar^2}{2m}d_{\ga}^x[d^x_{\ga}R\tl{E}_{\ga}+R\tl{E}_{\ga}z{\ga}^x]+VR
\tl{E}_{\ga}=
\end{equation}
$$=\frac{\hbar^2}{2m}\left[(d_{2\ga}^xR)\tl{E}_{\ga}+2d_{\ga}^xRz_{\ga}^x\tl{E}_{\ga}+R(z_{\ga}^x)^2\tl{E}_{\ga}+R\tl{E}_{\ga}z_{2\ga}^x\right]+VR\tl{E}_{\ga}$$
Then canceling the $\tl{E}_{\ga}$ terms in (2.31) gives
\bq\label{2.25}
i\hbar(R_t+Rz_t)=-\frac{\hbar^2}{2m}[d_{2\ga}^xR+2d^x_{\ga}Rz_{\ga}^x+R(z_{\ga}^x)^2+Rz_{2\ga}^x]+VR
\end{equation}
Using (2.23) now we have
\bq\label{2.26}
z_{\ga}^x=iz^{1-\ga}d_{\ga}^x\hat{S};\,\,z^{\ga}=
i\ga \hat{S};\,\, z^{\ga-1}z_{2\ga}^x=
d_{2\ga}^x \hat{S}-
\end{equation}
$$-(\ga-1)z^{\ga-2}\left[-z^{2(1-\ga)z}(d_{\ga}^x\hat{S})^2\right]
=id^x_{2\ga}\hat{S}+i\frac{(1-\ga)}{\ga}\frac{(d_{\ga}^x\hat{S})^2}{\hat{S}}$$
Thus we can use
\bq\label{2.27}
z_t=(\hat{S_t}/\hat{S})z;\,\,d_{\ga}^xz=z_{\ga}^x=iz^{1-\ga}d_{\ga}^x\hat{S};\,\,z^{\ga}=i\ga \hat{S}
\end{equation}
$$z^{(\ga-1)}z_{2\ga}^x=id^x_{2\ga}\hat{S}+i\frac{(1-\ga)}{\ga}\frac{(d_{\ga}^x\hat{S})^2}{\hat{S}}$$
(note that $z^{-\ga}=\ga/[\ga z^{\ga}]=i/\ga \hat{S}$).
\\[3mm]\indent
Now to determine how to split (2.25) we have to look at the complex numbers involved.  Thus $z^{\ga}/\ga=iS/\hbar$ and $z=i\hat{S}\ga$.  We assume $\hat{S}$ and $\ga$ are real with z complex so that
$({\bf 2L})\,\,z=[\hat{S}\ga]^{1/\ga}i^{1/\ga}=[\hat{S}\ga)]^{1/\ga}[exp(i\pi/2\ga]]$.  In this event we consider first the ingredients in (2.27) as complex numbers via the related z terms.  The terms in $d_{\ga},\,\,d_{2\ga}S,
\cdots$ are all real so one can most easily determine the form of the $z_{\ga}^x,\,\,z_{2\ga}^x,$ etc. from
(2.22)-(2.27).
\begin{enumerate}
\item
$z_t/z$ is real
\item
$z^{\ga-1}z_{\ga}^x$ is imaginary
\item
$z^{\ga-2}(z_{\ga}^x)^2=(1/z)(z^{\ga-1}z^x_{\ga})z^x_{\ga}$ is real (as the product of two imaginaries)
\item
$z^{\ga-1}z_{2\ga}^x$ is imaginary
\item
$z^{\ga}$ is imaginary
\end{enumerate}
Now refer to the classical situation where
\bq\label{2.28}
\hat{S}_t+\frac{\hat{S}_x^2}{2m}+V-\frac{\hbar^2 R''}{2m R}=0;\,\,\pp_tR+\frac{1}{m}(2R'\hat{S}'+R\hat{S}'')=0
\end{equation}
and the quantum potential (QP) is $({\bf 2M})\,\,{\mf Q}=-[\hbar^2R''/2mR]$.
Thus the correct choice from (2.24)-(2.25) appears to be
\bq\label{2.29}
i R_t=-\frac{1}{2m}\left[2d^x_{\ga}Rz_{\ga}^x+Rz^x_{2\ga}\right];\,\,i Rz_t=\frac{1}{2m}\left[(d_{2\ga}^xR+R(z_{\ga}^x)^2+VR\right]
\end{equation}
and we suggest a fractional QP
\bq\label{2.30}
{\mf Q}_{\ga}=-\frac{\hbar^2d^x_{2\ga}R}{2mR}
\end{equation}
which, not too surprisingly, agrees with \cite{cola}.  $\bs$
\\[3mm]\indent
Now the SE (2.15) of Jummarie was based on $D_x^{\ga}$ but conceals the information when expressed as 
in (2.15) in terms of $\pp_t$ and $\pp_x$.  In particular it makes it clumsy to identify a QP.  Hence we will check now a SE with $D^{\ga}$ for MRL of the type in (2.16) which immediately produces a QP
(cf. also \cite{hrmn} for SE of type (2.16)).
Thus for $\psi=Rexp(iS/\hbar)$ we write as usual
\bq\label{2.31}
D^{\ga}_x f(x)=f^{(\ga)}=\frac{d^{\ga}f}{(dx)^{\ga}}
\end{equation}
with the notation of Section 1.  Then consider a SE of classical form
\bq\label{2.32}
i\hbar\pp_t\psi=-\frac{\hbar^2}{2m}(D^{2\ga}_x\psi)+V\psi
\end{equation}
for $\psi=Rexp(iS/\hbar)$ which gives rise to the standard equations for $D_x^{2\ga}\to \gD$ (cf. \cite{c067}).  Recall that 
$D^{\ga}$
satisfies a chain rule and we look first for coarse grained space and classical time in writing
\bq\label{2.33}
D^{\ga}\left[(Re^{iS/\hbar})\right]=D^{\ga}Re^{iS/\hbar}+\frac{i}{\hbar}D^{\ga}RD^{\ga}Se^{iS/\hbar};
\end{equation}
$$D^{2\ga}\left[(Re^{\S/\hbar}\right]=D^{2\ga}Re^{iS/\hbar}+\frac{2i}{\hbar}D^{\ga}Re^{iS/\hbar}+$$
$$+R\left(\frac{i}{\hbar}\right)^2\frac{D^{2\ga}S}{\hbar}e^{iS/\hbar}+R\left(\frac{iD^{\ga}S}{\hbar}\right)^2
e^{iS/\hbar}$$
leading to (via (2.32))
\bq\label{2.34}
i\hbar\left(\pp_tR+\frac{iRS_t}{\hbar}\right)=-\frac{\hbar^2}{2m}\left[D^{2\ga}R+\frac{2iD^{\ga}RD^{\ga}S}{\hbar}-\frac{R}{\hbar^2}D^{2\ga}S-\frac{R(D^{\ga}S)^2}{\hbar^2}\right]
\end{equation}
which becomes
\bq\label{2.35}
i\hbar R_t-RS_t=-\frac{\hbar^2}{2m}D^{2\ga} R-\frac{i\hbar}{m}D^{\ga}RD^{\ga}S-\frac{R}{2m}D^{2\ga}S-
\frac{R}{2m}(D^{\ga}S)^2+\frac{VR}{m}
\end{equation}
and finally
\bq\label{2.36}
\pp_tR^2=\frac{1}{m}D^{\ga}(R^2)D^{\ga}S;\,\,-RS_t=-\frac{\hbar^2}{2m}D^{2\ga}R-\frac{1}{2m}(D^{\ga}S)^2
-\frac{V}{m}
\end{equation}
with QP
\bq\label{2.37}
{\mf Q}^{\ga}=-\frac{\hbar^2D^{2\ga}R}{2mR}
\end{equation}
as in (2.30).  Evidently this form will persist for any fractional derivative with a chain rule and Leibnitz
rule for products.  If in addition $t$ is coarse grained with a $D^{\gb}$ derivative then the $i\hbar\pp_t\psi$
term in (2.37) becomes 
\bq\label{2.38}
i\hbar D^{\gb}_t(Re^{iS/\hbar}=i\hbar\left[(D^{\gb}R)e^{iS/\hbar}+R\frac{i}{\hbar}D^{\gb}_tS e^{iS/\hbar}\right]
\end{equation}
so in (2.36B) we can replace $\pp_t$ by $D^{\gb}$ and the QP is unaffected.  Indeed (2.38) becomes
\bq\label{2.39}
D^{\gb}_tR^2=\frac{1}{m}D^{\ga}(R^2)D^{\ga}S;\,\,-RD^{\gb}_tS=-\frac{\hbar^2}{2m}D^{2\ga}R-\frac{1}{2m}(D^{\ga}S)^2-\frac{V}{m}
\end{equation}
Such results will hold for any choice of fractional derivatives satisfying the chain rule and Leibnitz
product rules.  It is worth noting that the MRL derivatives are non-local and solutions
(if any) of fractional SE may be sparse or have rather different properties than in the traditional
situations (cf. \cite{bayn,jxhs} for example).

\section{Q-DEFORMED AND FRACTIONAL CALCULUS}
\renewcommand{\theequation}{3.\arabic{equation}}
\setcounter{equation}{0}

We extract here from \cite{hrmn} (1007.1084) and refer also to other citations in \cite{hrmn}
for more information (especially \cite{bnds}).  It can be shown that the concept of q-deformed Lie algebras and the methods
developed in fractional calculus are closely related and may be combined leading to a new
class of fractional q-deformed Lie algebras (cf. also \cite{c002} for q-calculus).  In order now
to describe a  deformed Lie algebra we introduce a parameter q and define a mapping
\bq\label{16.1}
[x]_q=\frac{q^x-q^{-x}}{q-q^{-1}};\,\,lim_{x\to 1}[x]_q=x;\,\,[0]_q=0
\end{equation}
\bq\label{16.2}
D_x^qf(x)=\frac{f(qx)-f(q^{-1}x)}{(q-q^{-1})x};\,\,D_x^qx^n=[n]_qx^{n-1}
\end{equation}
As an example of a q-deformed Lie algebra one looks at the harmonic oscillator.
The creation and annihilation operators $a^{\dg},\,\,a$ and the number operator n
generate the algebra
\bq\label{16.3}
[N,a^{\dg}]=a^{\dg};\,\,[N,a]=-a;\,\,aa^{\dg}-q^{\pm 1}a^{\dg}a=q^{\mp N}
\end{equation}
Via (3.1) and alternative definition for (3.3) is given via $(\bullet)\,\,a^{\dg}a=[N]_q$ with
$aa^{\dg}=[N+1]_q$.  One defines a vacuum state with $a|0>=0$ and the action of the operators $\{a,a^{\dg},N\}$ on the basis $|n>$ of a Fock space is given by
\bq\label{16.4}
N|n>=n|n>;\,\,a^{\dg}|n>=\sqrt{[n+1]_q}\,|n+1>;\,\,a|n>=\sqrt{[n]_q}|n-1>
\end{equation}
The Hamiltonian of the q-deformed harmonic oscillator and its eigenvectors are defined via
\bq\label{16.5}
H=\frac{\hbar\go}{2}(aa^{\dg}+a^{\dg}a);\,\,E^q(n)=\frac{\hbar\go}{2}([n]_q+[n+1]_q)
\end{equation}
In \cite{hrmn} various fractional derivatives are recalled (cf. Sections 1-2) and in particular one
mentions the Caputo derivative
\bq\label{16.6}
D_x^{\ga}=\left\{\begin{array}{cc}
\frac{1}{\gG(1-\ga)}\int_0^xdx(x-\xi)^{-\ga}\pp_{\xi}f(\xi) & 0\leq \ga<1\\
\frac{1}{\gG(2-\ga)}\int_0^xd\xi(x-\xi)^{1-\ga}(\pp^2f(\xi)/\pp \xi^2) & 1\leq \ga<2
\end{array}\right.
\end{equation}
Then for $x^{n\ga}$ 
\bq\label{16.7}
D_x^{\ga}x^{n\ga}=\left\{\begin{array}{cc}
\frac{\gG(1+n\ga)x^{(n-1)\ga}}{\gG(1+(n-1)\ga)} & n>0\\
0 & n=0
\end{array}\right.
\end{equation}
The fractional derivative parameter $\ga$ can be interpreted as a deformation
parameter via $|n>=x^{n\ga}$ and
\bq\label{16,8}
[n]_{\ga}|n>=\left\{\begin{array}{cc}
\frac{\gG(1+n\ga)}{\gG(1+(n-1)\ga)}|n> & n>0\\
0 & n=0
\end{array}\right.;\,\,lim_{\ga\to 1}[n]_{\ga}=n
\end{equation}
Then via (3.2) the standard q-numbers can be defined more or less heuristically and
there are different possibilities.  On the other hand the q-deformation based on a
fractional calculus $\ga$ is uniquely determined once a set of basis vectors is given
and the harmonic oscillator will be used as an illustration.  This means that information about
q-entropy or q-information can be connected to any background $\ga-$fractional derivatives.
\\[3mm]\indent
Now replace $x$ and $p$ by $\hat{x}$ and $\hat{p}$ to get into QM and write
\bq\label{16.9}
\hat{X}f(x)=xf(x);\,\,\hat{P}f(x)=-i\hbar\pp_xf(x);\,\,[\hat{X},\hat{P}]=i\hbar
\end{equation}
Using $D_x^{\ga}$ there follows now
\bq\label{16.10}
\hat{x}=\left(\frac{\hbar}{mc}\right)^{1-\ga}x^{\ga};\,\,\hat{p}=-i\left(\frac{\hbar}{mc}\right)^{\ga}
mcD_x^{\ga}
\end{equation}
The classical and quantum Hamiltonians are now
\bq\label{16.11}
H_{class}=\frac{p^2}{2m}+\frac{1}{2}m\go^2x^2;\,\,H^{\ga}=\frac{\hat{p}^2}{2m}+\frac{1}{2}
m\go^2\hat{x}^2
\end{equation}
and the SE becomes
\bq\label{16.12}
H^{\ga}\psi=\left[-\frac{1}{2m}\left(\frac{\hbar}{mc}\right)^2m^2 c^2D_x^{\ga}D_x^{\ga}
+\frac{1}{2}m\go^2\left(\frac{\hbar}{mc}\right)^{2(1-\ga)}x^{2\ga}\right]\psi=E\psi
\end{equation}
The ``Hermiticity" of such an operator will depend of course on the choice of fractional
derivative and it can be shown that the Feller and Riesz fractional derivatives (but not
Caputo or Riemann-Liouville) will insure Hermiticity (cf. \cite{lask,hrmn}).  Here the Riesz
derivative is 
\bq\label{3.13}
D_R^{\ga}f(x)=\gG(1+\ga)\frac{Sin(\pi\ga/2)}{\pi}\int_0^{\infty}\frac{f(x+\xi)-2f(x)+f(x-\xi)}{\xi^{\ga+1}}d\xi;\,\,
(0<\ga<2)
\end{equation}
and the Feller derivative is 
\bq\label{3.14}
{}_FD_1^{\ga}f(x)=\gG(1+\ga)\frac{Cos(\pi\ga/2)}{\pi}\int_0^{\infty}\frac{f(x+\xi)-f(x-\xi)}{\xi^{\ga+1}}d\xi;\,\,(0\leq \ga<1)
\end{equation}
For a canonical
picture a scaled energy $E^q$ and coordinates can be introduced via
\bq\label{3.15}
\xi^{\ga}=\sqrt{\frac{m\go}{\hbar}}\left(\frac{\hbar}{mc}\right)^{1-\ga}x^{\ga};\,\,E=\hbar\go E^{\ga}
\end{equation} 
leading to the Hamiltonian
\bq\label{3.16}
H^{\ga}\psi_n(\xi)=\frac{1}{2}\left[-D^{2\ga}_{\xi}+\xi^{2\ga}\right]\psi_n(\xi)=E'(n,\ga)\psi_n(\xi)
\end{equation}
Laskin (cf. \cite{lask}) has derived an approximate analytic solution within the framework
of the WKB approximation which has the advantage of being independent of the choice
of a specific definition of the fractional derivatives (cf. \cite{hrmn,lask} for more information on
this) and the result is
\bq\label{16.17}
E'(n,\ga)=\left[\frac{1}{2}+n\right]^{\ga}\pi^{\ga/2}\left[\frac{\ga\gG\left(\frac{1+\ga}{2\ga}\right)}{\gG(1/2\ga)}\right];\,\,n=0,1,2,\cdots
\end{equation}
(cf. also \cite{gwhn,lask,trsv,wgch}).  We write also for a harmonic oscillator (following \cite{hrmn})
the connection to q-deformation arises from $n>\sim x^{n\ga}$ with $(\bl)\,\,E^q(n)=(\hbar\go/2)
([n]_q+[n+1]_q)$ while
\bq\label{3.18}
[n]_{\ga}|n>=\frac{\gG(1+n\ga)}{\gG(1+(n-1)\ga)}|n>;\,\,(\ga\sim q)
\end{equation}
for $n>0$ (as in (3.7)-(3.8)).  The q-deformation is then uniquely defined by $\ga$.

\section{THE Q EXPONENTIAL FAMILY}
\renewcommand{\theequation}{4.\arabic{equation}}
\setcounter{equation}{0}

We now begin to mix TD with the q-theory and from Naudts \cite{ndts} (cond-mat 0911.5392) we recall first the Boltzmann-Gibbs
distribution
\bq\label{4.1}
f_{\gb}(x)=\frac{c(x)}{Z(\gb)}e^{-\gb H(x)};\,\,Z(\gb)=\int dx c(x)e^{-\gb H(x)}
\end{equation}
Writing $\Phi(\gb)=log(Z(\gb))$ for the Massieu function there results
\bq\label{17.2}
\frac{d\Phi}{d\gb}=-U;\,\,U=E_{\gb}H=\int dx f_{\gb}(x)H(x);\,\,\frac{d^2\Phi}{d\gb^2}=
E_{\gb}(H-U)^2\geq 0
\end{equation}
Thus $\Phi(x)$ is convex as expected from TD and in fact $\Phi$ is by definition the
Legendre transform of the TD entropy $S(U)$ which is a convex function of U.  The
most common entropy functional is that of Boltzmann-Gibs-Shannon 
\bq\label{17.3}
I(f)=-\int dx f(x)log\frac{f(x)}{c(x)}
\end{equation}
One knows that $I(f)$ takes its maximum at $f=f_{\gb}$ over the set of all f with
$E_fH=U$.  This maximal value of $I(f)$ is then identified with the TD entropy
$S(U)$ via
\bq\label{17.4}
S(U)=max\{I(f);\,\,E_fH=U\}=I(f_{\gb})=
\end{equation}
$$=\int dx f_{\gb}(x)[log(Z(\gb))+\gb H(x)]=
log[Z(\gb))+\gb U$$
Consequently
\bq\label{17.5}
\Phi(\gb)=max_U\{S(U)-\gb U\}\Rightarrow S(U)=inf_{\gb}[\Phi(\gb)+\gb U]\Rightarrow
\gb=\frac{dS}{dU}
\end{equation}
Another consequence of (4.5) is that $S(U)$ is convex.
\\[3mm]\indent
The Boltzmann-Gibbs distribution defines a statistical model with
parameter $\gb$ and it belongs to the so called exponential family.
In fact a model belongs to the exponential family with parameters $\gt=
\{\gt_1,\cdots,\gt_n\}$ when its pdf $f_{\gt}(x)$ can be written in the
canonical form
\bq\label{17.6}
f_{\gt}(x)=c(x)exp\left[\sum_1^n\gt_jK_j(x)-\ga(\gt)\right]
\end{equation}
Models of this family share a number of properties.  For example they
share identities generalizing (4.2)
\bq\label{17.7}
\frac{\pp\ga}{\pp \gt_j}=E_{\gt}K_j=\int dx f_{\gt}(x)K_j(x)
\end{equation}
(the $K_j$ may not depend on $\gt$).
We recall now the q-deformed logarithm and exponential
\bq\label{17.8}
log_q(u)=\frac{1}{1-q}(u^{q-1}-1),\,\,(u>0);\,\,\frac{d}{du}log_q(u)=\frac{1}{u^q}>0
\end{equation}
The inverse function is 
\bq\label{17.9}
exp_q(u)=[1+(1-q)u]_{+}^{1/(1-q)}
\end{equation}
Then $({\bf 4A})\,\,exp_q[log_q(u)]=u;\,\,(u>0)$.  Further $exp_q(0)=1$
and
\bq\label{17.10}
\frac{d}{du}exp_q(u)=[exp_q(u)]^q\geq 0
\end{equation}
Now the notion of exponential family can be generalized by saying that the 
1-parameter model $f_{\gt}(x)$ belongs to the q-exponential family if there 
exist functions $c(x),\,\,\ga(\gt),\,\,H(x)$ such that $({\bf 4B})\,\,f_{\gt}(x)
=c(x)exp_q[-\ga(\gt)-\gt H(x)]$.  In the limit $q=1$ this is the standard definition
with $\gt=\gb$.  Then introduce the q-deformed entropy functional
\bq\label{17.11}
I_q(f)=-\int dx f(x) log_q\left(\frac{f(x)}{c(x)}\right)=\frac{1}{1-q}\left[1-\int dx
c(x)\left(\frac{f(x)}{c(x)}\right)^{2-q}\right]
\end{equation}
Assume now that $0<q<2$ and one can show that the pdf $f=f_{\gt}(x)$ maximizes the quantity
\bq\label{17.12}
\frac{1}{2-q}I_q(f)-\gt E_fH
\end{equation}
For $q=1$ this is known as a variational principle (cf. \cite{ndts}) which implies
the maximum entropy principle $I_q(f_{\gt})\geq I_q(f)$ for all $f$ which satisfy
$E_fH=E_{\gt}H$ (using the notation $E_{\gt}=E_{f_{\gt}}$.  A well-known problem is now that $f_{\gt}$ also maximizes $\xi(I_q(f))$ where $\xi(u)$ is an arbitrary monotonically increasing function.  Therefore a meaningful Ansatz is to assume
that the TD entropy $S(U)$ is given via
\bq\label{17.13}
S(U)=\xi(I_q(f_{\gt})=\xi(\ga(\gt)+\gt E_{\phi}H)
\end{equation}
The TD expression for the inverse temperature $\gb$ can be then calculated
using (4.5).  Evidently the resulting relaton between energy U and temperature
$\gb^{-1}$ will depend on the choice of the monotonic function $\xi(u)$.
\\[3mm]\indent
The q-Gaussian distribution in one variable is
\bq\label{17.14}
f(x)=\frac{1}{c_q\gs}exp_q(-x^2/\gs^2)
\end{equation}
with
\bq\label{17.15}
c_q=\int_{-\infty}^{\infty}dx exp_q(-x^2)=\sqrt{\frac{\pi}{q-1}}\frac{\gG(-\frac{1}{2}
+\frac{1}{q-1})}{\gG\left(\frac{1}{q^{-1}}\right)}\,\,(1<q<3)
\end{equation}
$$=\sqrt{\frac{\pi}{1-q}}\frac{\gG\left(1+\frac{1}{1-q}\right)}{\gG\left(\frac{3}{2}
+\frac{1}{1-q}\right)}\,\,(q<1)$$
It can be brought into the form ({\bf 4B}) with $c(x)=1/c_q,\,\,H(x)=x^2,\,\,\gt=\gs^{3-q}$ and
\bq\label{17.16}
\ga(\gt)=\frac{\gs^{q-1}}{q-1}=log_{2-q}(\gs)
\end{equation}
The $q=1$ case reproduces the congenital Gauss distribution and for $q<1$
the distribution vanishes outside an interval.  Take for instance $q=1/2$
in which case (4.14) becomes
\bq\label{17.17}
f(x)=\frac{15\sqrt{2}}{32\gs}\left[1-\frac{x^2}{\gs^2}\right]_{+}^2
\end{equation}
This distribution vanishes outside the interval $[-\gs,\gs]$ and in the range
$q<3$ the q-Gaussian is strictly positive on the whole line and decays
with a power law in $|x|$ instead of exponentially.  For $q=2$ there results
$({\bf 4C})\,\,f(x)=(1/\pi)(\gs/(x^2+\gs^2)$.
\\[3mm]\indent
Apparently there is no consensus about the correct definition of the TD
entropy $S(U)$ for isolated systems; the matter is important since it
directly determines the definition of the TD temperature via (4.5).  
Most often $({\bf 4D})\,\,S(U)=k_Blog[\go(U)]$ is used where $\go(U)$
is the density of states but this has drawbacks (cf. \cite{ndts}).
The shortcomings of Boltzmann's entropy have been noticed since long ago
and a slightly different definition of entropy is $({\bf 4E})\,\,S(U)=k_Blog[\gO(U)]$
where $\gO(U)$ is the integral $({\bf 4F})\,\,\gO(U)=(1/\hbar^{3N})\int \prod
d{\bf p}_i\int\prod d{\bf q}_i\gT(U-H({\bf q},{\bf p})$.  An immediate advantage of
({\bf 4E}) is that the resulting expression for the temperature T defined by $({\bf 4G})\,\,(1/T)=dS/dU$ coincides with the notion of temperature used by the experimentalists; indeed $({\bf 4H})\,\,k_BT=\gO(U)/\go(U)$.
For a harmonic oscillator the density of states $\go(U)$ is a constant so ({\bf 4H}) implies $k_BT=U$ as desired (cf. \cite{ndts} for more discussion).

\section{TD AND FRACTALS}
\renewcommand{\theequation}{5.\arabic{equation}}
\setcounter{equation}{0}

It seems presumptive to simply fractalize an arbitrary intensive or extensive TD variable in order to see
what happens.  But it does seem possible to ``sneak" fractals into TD via entropy ideas and fractional
calculus.  For more on this theme we sketch from \cite{ezek,ezgb}
Thus the purpose in \cite{ezek} is to point out that q-derivatives are naturally suited for describing systems
with discrete dilatation symmetries such as fractal and multi-fractal sets, where the limit $q\to 1$ corresponds
to continuous scale change.  Thus the q-derivative can be defined via
\bq\label{14.1}
\pp_x^{(q)}f(x,y,\cdots)=\frac{f(qx,y,\cdots)-f(x,y,\cdots)}{(q-1)x}
\end{equation}
This derivative measures the rate of change with respect to a dilatation of its
arguments by a factor of $q$.  Even for some functions with no ordinary derivative  a sequence of $q's$ going to 1 can be chosen in such a manner that the limit is well defined.  We recall that self-similar
sets are characterized by homogeneous functions, e.g. $f$ is of degree $\psi$ means
\bq\label{14.2}
\pp_x^{(q)}f(x)=\frac{q^{\psi}-1}{q-1}\frac{f(x)}{x}
\end{equation}
Following standard notation the q-number 
\bq\label{14.3}
[\psi]_q=\frac{q^{\psi}-1}{q-1}\Rightarrow x\pp_x^{(q)}f(x)=[\psi]_qf(x)
\end{equation}
It is useful to note that the solution of $({\bf 5A})\,\,\pp_x^{(q)}f(x)=0$ is either a constant or a function
periodic in $log(x)$, with period $log(q)$, such that $A_q(qx)=A_q(x)$.  With the modified product rule
\bq\label{14.4}
\pp_x^{(q)}[g(x)f(x)]=g(qx)\pp_x^{(q)}f(x)+f(x)\pp_x^{(q)}g(x)
\end{equation}
there is, besides the artitrary additive constant of integration, an arbitrary multiplicative function of integration
satisfying $({\bf 5B})\,\,f(x)=A_q(x)x^{\psi}$ (taking the constant of integration to be 0).  The function
$A_q(x)$ has some TD meaning, e.g. the free energy has the form ({\bf 5B}) and can be expressed
in the form of a q-integral (cf. \cite{eezn}).  For $0<q<1$ the q-integral is defined as
\bq\label{14.5}
\int_0^xD_t^{(q)}g(t)=(1-q)x\sum_0^{\infty}q^ng(q^nx);\,\,\int_x^{\infty}D_t^{(q)}g(t)=(1-q)\sum_1^{\infty}q^{-n}g(q^{-n}x)
\end{equation}
where $D_t^{(q)}$ denotes the q-differential.  For any given $\psi>0$ and $qne 1$ a function $A_q(x)$ with the desired property ({\bf 5A}) can be written via
\bq\label{14.6}
A_q(x)=x^{-\psi}\sum_{-\infty}^{\infty}\frac{g(q^nx)}{q^{n\psi}}=
\end{equation}
$$=\frac{1}{1-q}\int_0^x\frac{g(t)}{t^{1+\psi}}
D_t^{(q)}+\frac{1}{1-q}\int_x^{\infty}\frac{g(t)}{t^{1+\psi}}D_t^{(q)}$$
The function $g(x)$ is quite arbitrary (see \cite{ezek} for more details).  An example here is
the Weierstrass-Mandelbrot function
\bq\label{14.7}
f_b(x)=\sum_{-\infty}^{\infty}b^{n\psi}[1-Cos(b^nx)];\,\,b>1,\,\,0<\psi<1
\end{equation}
The limit 
\bq\label{14.8}
lim_{b\to 1}\frac{x\pp_x^{(b)}f_b(x)}{f_b(x)}=\psi
\end{equation}
but $lim_{b\to 1}xf'_b(x)/f_b(x)$ is not defined.
\\[3mm]\indent
Now consider a fractal set S which is self-similar under scale changes by $b$ and let $\rho_b(x)$ satisfy
({\bf 5A}) with $q=b$ (with respect to any one of its arguments $x_k,\,\,k=1,\cdots, d$ where d is the
dimension of the embedding space.  This function can be constructed as the $n\to\infty$ limit of a succession of $\rho_b^{(n)}$ such that
\bq\label{14.9}
\rho_b^{(n)}(x)=\left\{
\begin{array}{cc}
1 & x\in \cup_i\,\gO_i(b^n)\,\,(i=1,\cdots,N_n)\\
0 & otherwise
\end{array}\right.
\end{equation}
where $\gO_i(b^n)$ is a covering of the set S consisting of $N_n=N^n$ balls of linear size $b^n$ ($b<1$ and
$N>1$ is arbitrary).  The fractal self-similarity dimension of this set is $d_f=log(N)/log(b^{-1})$ and to see
this define
\bq\label{14.10}
M(\ell_n,R)=\int_{\gG(R)}\rho_b^{(n)}(x)d^dx=\Phi_b(\ell_n,R)\ell^d_n(\ell_n/R)^{-d_f}
\end{equation}
where $\ell_n=b^n\ell_0$.  $\gG(R)$ is a region of linear size R of the embedding space of dimension d.
The oscillatory amplitude $\Phi_b(\ell,R)$ is called lacunarity and is periodic in the logarithm of both $\ell$ and R, with period $log(b)$.  As an example consider the Cantor set with $d=1,\,\,\ell_0=1,\,\,b=1/3,$ and
$N=2$.  From (5.10) it is clear that $M(\ell_n,1/3)/M(\ell_n,1)=1/2=(1/3)^{d_f}$ yielding $d_f=log(2)/og(3)$.
Now taking the limit of $n\to\infty$ one can define $({\bf 5D})\,\,M(R)=lim_{n\to\infty}M(\ell_n,R)=\int_{\gG(R)}\rho_b(x)d^dx$.  The usefulness of this definition is that $M(R)$ is an eigenfunction of the operator
$R\pp_R^{(q)}$ with $q=b$ being the scale change which takes the fractal set to itself such that
$({\bf 5E})\,\,R\pp_R^{(b)}M(R)=[d_f]_bM(R)$, while its derivative fails to exist at an infinite number of points.
It is convenient to introduce the dilatation operator such that $T_x^{(q)}f(x)=f(qx)$.  Then $T_x^{(q)}=[1+(q-1)x\pp_x^{(q)}]$ or $T_x^{)q)}=[\pp_x^{(q)},x]$ where the brackets indicate the ordinary commutator.
Then S is generated by an exact recursion relation based on a fine graining of the covering by $b$, leading
to $({\bf 5F})\,\,T_{\ell}^{(b)}M(\ell_n,R)=M(\ell_{n+1},R)$ or $({\bf 5G})\,\,(T^{(b)}_{\ell})^kM(\ell,R)=M(b^{\ell}\ell,R)$.  For small $b-1$ exponentiation yields 
\bq\label{14.11}
M(b^{\ell}\ell,R)=(T_{\ell}^{(b)})^kM (\ell,R)= e^{k[T_{\ell}^{(b)}-1]}M(\ell,R)
\end{equation}
There is also a discussion of multi-fractal situations with considerable detail.
\\[3mm]\indent
We go now briefly to \cite{masi} for further connections of generalized entropies to Fisher information
(with some repetition - cf. also \cite{frch}).  First from \cite{masi} note that the Renyi entropy
isn't convex and does not have the property of finite entropy production, so no extension can have
these properties either.  There is some controversy about TD implications here; but this does not
affect the meaning and applications in e.g. information theory.  First write the Shannon entropy
in the form  $({\bf 5H})\,\,S_S(P)=-\sum_ip_ilog(p_i)=\sum p_i log(1/p_i)$ while the Tsallis entropy
is
\bq\label{14.12} 
S_T(P,q)=\frac{\sum_ip_i^q-1}{1-q}=\frac{1}{q-1}\sum p_i\left[1-p_i^{q-1}\right]
\end{equation}
It is convenient here to use the notation
\bq\label{14.13}
log_q(x)=\frac{x^{1-q}-1}{1-q};\,\,e_q^x=[1+(1-q)x]^{1/(1-q)};
\end{equation}
$$log_q(xy)=log_qx+log_qy+(1-q)(log_q(x))
(log_q(y))$$
Exploiting this general notation the Tsallis entropy is  $({\bf 5I})\,\,S_T(P,q)=-\sum p_i^qlog_q(p_i)=\sum p_ilog_q[1/p_i]$ which is sometimes called a q-deformed Shannon entropy.  Then define an information measure via 
\bq\label{14.14}
I_i=I_i\left(\frac{1}{p_i}\right)=log_q\left(\frac{1}{p_i}\right);\,\,S_S(P)=\left<\log\left(\frac{1}{p_i}\right)\right>_{lin};
\end{equation}
$$S_T(P)=\left<log_q\left(\frac{1}{p_i}\right)\right>_{lin}$$
More generally, following Kolmogorov and Nagumo one can write $({\bf 5J})\,\,S=f^{-1}\left(\sum_ip_if(I_i)\right)$ with $f$ a strictly monotone continuous function (KN function).  Renyi instead showed that, if
additivity is imposed on information measures, then the whole set of KN functions must reduce to two cases,
namely $({\bf 5K})\,\,f(x)=x$ or $f(x)=c_1b^{(1-q)x}+c_2$.  Renyi's information-entropy measure is then
\bq\label{14.15}
S_R(P,q)=\frac{1}{1-q}log_b\sum_ip_i^q=\frac{1}{1-q}log\sum_ip_i^q
\end{equation}
(assume here $b=e$ is the logarithm base).  In fact if one chooses in ({\bf 5K}) $c_1=1/(1-q)-c_2$ then
there follows 
\bq\label{14.16}
f(x)=log_qe^x;\,\,I_i=log\left(\frac{1}{p_i}\right);\,\,S_R(P,q)=\left<log\left(\frac{1}{p_i}\right)\right>_{exp}
\end{equation}
(where $<\,\,\,>$ is the KN average in (5.16)).
\\[3mm]\indent
Now introduce the Sharma-Mittal and Supra-extensive entropies, beginning with the relation between
Tsallis and Renyi entropies
\bq\label{14.17}
S_R(P,q)=\frac{1}{1-q}log[1+(1-q)S_T(P,q)]
\end{equation}
This is equivalent to
\bq\label{14.18}
S_R(P,q)=log\left[e_q^{S_T(P,q)}\right]\Rightarrow S_T(P,q)=log_q\left[e^{S_R(P,q)}\right]
\end{equation}
This suggests two possible further generalizations
\bq\label{14.19}
S_{SM}(P,\{q,r\})=log_re_q^{S_T(P,q)}=\frac{1}{1-r}\left[\left(\sum_ip_i^q\right)^{\frac{1-r}{1-q}}-1\right]
\end{equation}
\bq\label{14.20}
S_{SE}(P,\{q,r\})=log_qe_r^{S_R(P,q)}=\frac{\left[1+\frac{1-r}{1-q}log\sum_ip_i^q\right]^{\frac{1-r}{1-q}}-1}{1-q}
\end{equation}
(cf. \cite{masi} for more on this).  Some rewriting is also possible via
\bq\label{14.21}
\left(\sum_ip_i^q\right)^{1/(1-q)}\left(\sum_ip_i\left(\frac{1}{p_i}\right)^{1-q}\right)^{1/(1-q)}=
\end{equation}
$$\left<\left(\frac{1}{p_i}\right)^{1-q}\right>_{lin}^{1/(1-q)}=e_q^{<log_q(1/p_i)>_{lin}}
=e_q^{S_T(P,q)}=\left<\frac{1}{p_i}\right>_{log_q}=\gO(P,q)$$
where the logarithmic mean $<\cdot>_{log_q}\equiv\gO(P,q)$ is used.  One can then rewrite (5.21)
in the form
\bq\label{14.22}
S_T(P,q)=log_q\gO(P,q);\,\,S_R(P,q)=log\gO(P,q);
\end{equation}
$$S_{SM}(P,(q,r))=log_r\gO(P,q);\,\,S_{SE}(P,(q,r))=
log_qe_r^{\gO(P,q)}$$
\bq\label{14.23}
\gO(P,q)=\left<\frac{1}{p_i}\right>_{log_q}=(\sum p_i^q)^{1/(1-q)}
\end{equation}

\indent
Recall also the Kullback-Leibler relative information-entropy measure ($SP\sim$ sample space)
\bq\label{14.24}
S_{KL}=\int_{SP}p_1(x,\gt)\log\frac{p_2(x,\phi)}{p_1(x,\gt)}d^nx
\end{equation}
and define then the information measures of Tsallis, Renyi, etc. as
\bq\label{14.25}
S_T({\mc P},q)=\frac{1}{1-q}\left[\int p_1(x,\gt)^qp_2(x,\phi)^{1-q}d^nx-1\right]
\end{equation}
$$S_R({\mc P},q)=\frac{1}{1-q}log\int p_1(x,\gt)^qp_2(x,\phi)^{1-q}d^nx;$$
$$S_{SM}({\mc P}(q,r))=\frac{1}{1-r}\left[\left(\int p_1(x,\gt)^qp_2(x,\phi)^{1-q}d^nx\right)^{\frac{1-r}{1-q}}-1\right]$$
$$S_{SE}({\mc P},(q,r))=\frac{\left[1+\frac{1-r}{1-q}log\int p_1(x,\gt)^qp_2(x,\phi)^{1-q}\right]^{\frac{1-r}{1-q}}-1}{1-q}$$
The Fisher information $I_F$ is defined via
\bq\label{14.26}
I_F({\mc P})=\int\frac{1}{p(x,\gt)}\left(\frac{\pp p(x,\gt)}{\pp \gt}\right)^2dx
\end{equation}
and for any other estimator one has a Cramer-Rao inequality $({\bf 5L})\,\,I_Fe^2\geq 1$.
For n-dimensions the Fisher information matrix is defined via
\bq\label{14.27}
F_{ij}(\gt)=\int\frac{1}{p(x,\gt)}\frac{\pp p(x,\gt)}{\pp \gt_i}\frac{\pp p(x,\gt)}{\pp\gt_j}d^nx
\end{equation}
or as an expectation value $({\bf 5M})\,\,F_{ij}(\gt)=E(\pp_i\ell(\gt)\pp_j\ell(\gt)]$ where $(\ell\sim$
partial derivatives).  Note that symmetrization is needed by means of $({\bf 5N})\,\,{\mc D}(p_1,p_2)=
\frac{1}{2}\left[S(p_1,p_1)+S(p_2,p_1)]\right]$ to obtain
\bq\label{14.28}
g_{ij}(\gt,\phi)=\frac{1}{2}\left[\frac{\pp^2(S(\gt,\phi)+S(\phi,\gt))}{\pp \gt_i\pp\gt_j}\right]_{\phi=\gt}
\end{equation}
Then write also $({\bf 5O})\,\,{\mc D}(\gt,\phi)=(1/2)[S(\gt,\phi)+S(\phi,\gt)]$ and it results that
$({\bf 5P})\,\,g_{ij}^{SE}(\gt)=g_{ij}^{SM}(\gt)=g_{ij}^R(\gt)=g^T_{ij}(\gt)=qg_{KL}(\gt)=-qF_{ij}(\gt)$.
\\[3mm]\indent
Next one establishes some relations between the various entropies mentioned above and the
Fisher information.  Thus set $({\bf 5Q})\,\,\pp_xlog_q(x)=1/x^q$ and $\pp_xe_q^x=(e_q^x)^q$.  Then
\bq\label{14.29}
S_T(\gt,\phi)=\int p_1(x,\gt)log_q\left(\frac{p_2(x,\phi)}{p_1(x,\gt)}\right)d^nx;
\end{equation}
$$S_T(\phi,\gt)=
\int p_2(x,\phi)log_q\left(\frac{p_1(x,\gt)}{p_2(x,\phi)}\right)d^nx$$
The resulting $(ij)$ derivatives are quite different than one would expect from (5.25)
with identification only for normalized PDF.  For distributions as in ({\bf 5N}) there is some simplification
however and 
\bq\label{14.30}
[\pp_{ij}S_T(\gt,\phi)]_{\phi=\gt}=[\pp_{ij}S_T(\phi,\gt)]_{\phi=\gt}=
\end{equation}
$$=-q\int\frac{1}{p(x,\gt)}\pp_ip(x,\gt)\pp_jp(x,\gt)d^nx=-qF_{ij}(\gt)$$
which via (14.28) yields $g^T_{ij}=-qF_{ij}(\gt)$
Similarly one has 
\begin{enumerate}
\item
$({\bf 5R})\,\,[\pp_{ij}S_R]_{\phi=\gt}=[\pp_{ij}S_T]_{\phi=\gt}\Rightarrow
g_{ij}^R=g_{ij}^T=-qF_{ij}$
\item
$({\bf 5S})\,\,[\pp_{ij}S_{SM}]_{\phi=\gt}=[\pp_{ij}S_T]_{\phi=\gt}=-qF_{ij}$
\item
$({\bf 5T})\,\,g_{ij}^{SE}=g_{ij}S_T=-qF_{ij}$
\end{enumerate}
Thus the Fisher information accounts (modulo $-q$) for the change of multiplicity (or the number of
microstates) under a statistical parameter variation.

\section{Q-FISHER AND TSALLIS}
\renewcommand{\theequation}{6.\arabic{equation}}
\setcounter{equation}{0}

We follow here \cite{frch} (JMP 50 (2009), -13303) (cf. also \cite{akdt,bchr,cruz,eezn,ezek,frch,frmt,fryg,gwhn,
hpvv,hrvp,jbar,koel,lprz,lnrs,masi,pnpl,ppfr,pppi,syri,ubrc,wgch}.
Denote the q-logarithm and exponential via
\bq\label{15.1}
log_q(x)=\frac{x^{1-q}-1}{1-q};\,\,exp_q(x)=\left\{\begin{array}{cc}
[1+(1-q)x]^{1/(1-q)} & 1+(1-q)x>0\\
0 & otherwise
\end{array}\right.
\end{equation}
and note that
\bq\label{15.2}
exp_q[x+y+(1-q)xy]=exp_q(x)exp_q(y);\,\,
\end{equation}
$$log_q(xy)=log_q(x)+log_q(y)+(1-2)log_q(x)log_q*y)$$
The set of all probability densityfunctions on $({\bf R})$ is denoted by$({\bf 6A})\,\,D=\left\{f:\,{\bf R}\to {\bf R};\,\,f(x)\geq 0;\,\,\int_{-\infty}^{\infty}f(x)dx=1\right\}$ and the Tsallis relative entropy is $H_q(\phi(x))=-\int_{-\infty}^{\infty}\phi(x)^qlog_q\phi(s)dx$.  Thus
\bq\label{15.3}
D_q(\phi(x)|\psi(x))=\int_{-\infty}^{\infty}\phi(x)^q[log_q(\phi(x))-log_q(\psi(x))]dx
\end{equation}
For $q\to 1$ the Tsallis q-entropy converges to the Shannon entropy and the relative entropy converges to the Kullback-Leibler (KL) divergence $(\bullet)\,\,D_1(\phi(x)|(\psi(x))=\int_{-\infty}^{\infty}]\phi(x)[log(\phi(x))-log(\psi(x))dx$.  Define now two constraints on the normalized q-expectation value and q-variance, namely
\bq\label{15.4}
C_q^{(c)}=\left\{f\in D;\,\,\frac{1}{c_q}\int_{-\infty}^{\infty}xf(x)^qdx=\mu_q\right\};
\end{equation}
$$C_q^{(g}=\left\{f\in C_q^{(c)};\,\,\frac{1}{c_q}\int_{\infty}^{\infty}(x-\mu_q)^2f(x)^qdx=\gs_q^2\right\}$$
The q-canonical distribution $\phi_q^{(c)}(x)\in D$ and the q-Gaussian distribution $\phi_q^{(g)}\in D$
are
\bq\label{15.5}
\phi_q^{(c)}(x)=\frac{1}{Z_q^{(c)}}exp_q\left[-\gb_q^{(c)}(x-\mu_q)\right];\,\,Z_q^{(c)}=\int_{\infty}^{\infty}
exp_q\left[-\gb_q^{(c)}(x-\mu_q)\right];
\end{equation}
$$\phi_q^{(g)}(x)=\frac{1}{Z_q^{(g)}}exp_q\left[\frac{-\gb_q^{(g)}(x-\mu_q)^2}{\gs_q^2}\right];\,\,Z_q^{(g)}=
\int_{-\infty}^{\infty}exp_q\left[\frac{-\gb_q^{(g)}(x-\mu_q)^2}{\gs_q^2}\right]$$
One shows next that $({\bf 6B})\,\,D_q(\phi(x)|\psi(x))\geq 0$ with equality if and only if $\phi(x)=\psi(x)$ for
all $x$.  Further 
\bq\label{15.6}
H_q(\phi(x))\leq -c_q\frac{1}{Z_q^{(c)}};\,\,equality \iff \phi(x)=\frac{1}{Z_q^{(c)}}exp_q\left[-\gb_q^{(c)}(x-\mu_q)\right]
\end{equation}
(cf. \cite{frch} for proof).  The generalized free energy takes a minimum
\bq\label{15.7}
F_q=\mu_q-\frac{1}{\gb_q^{(c)}}H_q(\phi(x))\geq\mu_q+\frac{c_q}{\gb_q^{(c)}}log_q\frac{1}{Z_q^{(c)}}
\end{equation}
if and only if $\phi(x)=(1/Z_q^{(c)})exp_q[-\gb_q^{(c)}(x-\mu_q)]$.  Further if $\phi\in C_1^{(c)}$ then 
$H_1(\phi(x))\leq log(Z_1^{(c)}$ with equality $\iff\,\, \phi(x)=(1/Z_1^{(c)})exp[-\gb_1^{(c)}(x\mu)]$.
Next it can be shown that for $0\leq q<3$ and $q\ne 1$, if $\phi\in C_q^{(g)}$ 
\bq\label{15.8}
H_q(\phi(x))\leq -c_qlog_q\frac{1}{Z_q^{(g)}}+c_q\gb_q^{(g)}[Z_q^{(g)}]^{q-1}
\end{equation}
with equality if and only if
\bq\label{15.9}
\phi(x)=\frac{1}{Z_q^{(g)}}exp_q\left[-\frac{\gb_q^{(g)}(x-\mu_q)^2}{\gs_q^2}\right]
\end{equation}
where $Z_q^{(g)}=\int_{-\infty}^{\infty}exp_q\left[-\gb_q^{(g)}(x-\mu_q)^2/\gs_q^2)\right]dx$
with $\gb_q^{(g)}=1/(3-q)$.
\\[3mm]\indent
Now the fact that the Gaussian distribution minimizes the Fisher information leads to the study the Tsallis distribution (q-Gaussian) to see if it minimizes a q-Fisher information as a 1-parameter extension.
In this direction one defines the q-score function $s_q(x)$ and the q-Fisher information $J_q(x)$ via
\bq\label{15.10}
s_q(x)=\frac{d\,log_qf(x)}{dx};\,\,J_q(X)=E_q[s_q(x)^2]
\end{equation}
where $E_q$ is defined as $E_q[g(X)]=\frac{\int g(x)f(x)^qdx}{\int f(x)^qdx}$ for random variables
$g(X)$, with continuous $g(x)$ and a probability density $f(x)$ (note the definition of $J_q$ differs from
\cite{pppi} for example).  Then as an example for a random variable G obeying a q-Gaussian distribution
consider
\bq\label{15.11}
\phi_q^{(g)}(x)=\frac{1}{Z_q}exp_q\left[-\frac{\gb_q(x-\mu_q)^2}{\gs_q^2}\right]
\end{equation}
where $\gb_q=1/(3-q)$ and the q-partition function is $Z_q=\int_{-\infty}^{\infty}exp_q[(x-\mu_q)^2/\gs_q^2]
dx$, with q-score function and q-Fisher information 
\bq\label{15.12}
s_q(x)=-\frac{2\gb_qZ_q^{q-1}}{\gs_q^2}(x-\mu_q);\,\,J_q(G)=\frac{4\gb_q^2Z_q^{2q-2}}{\gs_q^2}
\end{equation}
so $lim_{q\to 1}J_q(G)=1/\gs_1^2$.  Then given the random variable X with probability density function
$p(x)$, the q-expectation value $\mu_q=E_q(X)$, and the q-variance $\gs_q^2=E_q(X-\mu_q)^2]$.  There is then a q-Cramer-Rao inequality
\bq\label{15.13}
J_q(X)\geq\frac{1}{\gs_q^2}\left[\frac{2}{\int p(x)^qdx}-1\right];\,\,q\in[0,1]\cup (1,3)\Rightarrow
J_q(X)\geq\frac{1}{\gs_q^2}\,\,(q\in(1,3)
\end{equation}
To see this assume $lim_{x\to\pm\infty}\,\,f(x)p(x)=0$ for any $q\geq 0$, any probability density $p$
and any smooth function $f$ suitably behaved at $\pm\infty$.  Then using $p$ in the score function
(6.10)
\bq\label{15.14}
E_q(X-\mu_q)s_q(x)=\frac{\int (x-\mu_q)p(x)^2s_q(x)dx}{\int p(x)^qdx}=\frac{\int(x-\mu_q)p'(x)dx}{\int p(x)^qdx}=\frac{-1}{p(x)^qdx}
\end{equation}
Note for $q<1$ there is no relation between $J_q(X)$ and $1/\gs_q^2$ beyond (6.13).  Thus it seems
that the Tsallis entropies and q-Fisher information make sense only for the case of $q\geq 1$ in the present
setting.
\\[3mm]\indent
{\bf REMARK 6.1.}
We mention here for example the papers \cite{bcpp,chpp,cupt,flpp,hrvp,klmk,kmpz,mppv,mtpl,pnpl,plts,ppsf,pppi,vgpl,vppo}, involving
q-entropy, q-information, q-TD, Fisher information, the Frieden-Soffer idea of extreme physical information,
the SE, non-equilibrium TD, Legendre structure, etc.  They seem to signal new portals between TD, QM,
and gravitational physics to go along with the cosmological input of Padmanabhan 
\cite{pada,padn,padh} and
others connecting gravity and entropy (see e.g. \cite{jacb,vlde} and cf. also \cite{ccgn,clnd,cnsc,cgot} for 
fractal structure).  We mention also a series of papers \cite{gsbk,skhv,skgb,sbgv} connecting macro
(TD) physics and micro (QM) physics in a very interesting mannerand would like to refer as well to some earlier work involving TD and QM in e.g. \cite{acis,c067,crar,caro,grps,gfps}.  $\bs$

\end{document}